\documentclass{emulateapj}
\usepackage{amssymb}
\usepackage{natbib}
\usepackage{graphicx}
\usepackage{epstopdf}
\newcommand{\msun}{\ensuremath{\rm M_\odot}}
\newcommand{\Ha}{\ensuremath{\rm H\alpha}}
\newcommand{\lya}{\ensuremath{\rm Ly\alpha}}
\newcommand{\kms}{km~s\ensuremath{^{-1}}}
\newcommand{\ztwo}{\ensuremath{z\sim2}}
\newcommand{\zthree}{\ensuremath{z\sim3}}

\begin{document}

\title{Filamentary Large Scale Structure Traced by Six \lya\ Blobs at  $z=2.3$}
\author{{\sc Dawn K. Erb}\altaffilmark{1}, {\sc Milan Bogosavljevi{\'c}}\altaffilmark{2}, and {\sc  Charles C. Steidel}\altaffilmark{3}}

\slugcomment{Accepted for publication in ApJL}

\shorttitle{FILAMENTARY STRUCTURE AND \lya\ BLOBS AT $z=2.3$}
\shortauthors{ERB, BOGOSAVLJEVI{\'C} AND STEIDEL}

\altaffiltext{1}{Department of Physics, University of Wisconsin Milwaukee, Milwaukee, WI, 53211; dawn@gravity.phys.uwm.edu}
\altaffiltext{2}{Astronomical Observatory, Volgina 7, 11060 Belgrade, Serbia; mbogosavljevic@aob.rs}
\altaffiltext{3}{Department of Astronomy, California Institute of Technology, MS 249--17, Pasadena, CA 91125; ccs@astro.caltech.edu}

\begin{abstract}
Extended nebulae of \lya\ emission (``\lya\ blobs") are known to be associated with overdense regions at high redshift.  Here we present six large \lya\ blobs in a previously known protocluster with galaxy overdensity $\delta \sim 7$ at $z=2.3$; this is the richest field of giant \lya\ blobs detected to date.  The blobs have linear sizes of $\gtrsim100$ kpc and \lya\ luminosities of $\sim10^{43}$ erg s$^{-1}$.  The positions of the blobs define two linear filaments with an extent of at least 12 comoving Mpc; these filaments intersect at the center of one of the blobs. Measurement of the position angles of the blobs indicates that five of the six are aligned with these filaments to within $\sim10^{\circ}$, suggesting a connection between the physical processes powering extended \lya\ emission and those driving structure on larger scales.
\end{abstract}


\section{Introduction}
Large, extended regions of \lya\ emission, also known as \lya\ blobs, have been most commonly found in overdense regions at redshifts \ztwo\ to \zthree.  These objects have sizes of $\sim50$--100 kpc, \lya\ luminosities of $\sim10^{43}$--$10^{44}$ erg s$^{-1}$, and, in contrast to the large \lya\ nebulae surrounding some high redshift radio galaxies (e.g.\ \citealt{msd+87}), do not have obvious sources for their strong emission. Since the first two such blobs were found in an overdensity at $z=3.1$ \citep{sas+00}, additional blobs have been found in the same region \citep{myh+04,myh+05}, in other dense regions \citep{ptf+04,pkdm08,mnm+09}, and, less frequently, in the field \citep{yzt+09,myh+11}.  Lower redshift ($z=0.8$) searches for \lya\ blobs have so far found none \citep{kwcw09}, suggesting that the blobs are a high redshift phenomenon.

While the blobs' preferential location in overdense environments indicates an association with massive galaxy formation, the mechanism which powers the \lya\ emission remains unclear.   Proposed sources have generally fallen into two categories: cooling radiation from cold streams of gas accreting onto galaxies (e.g.\ \citealt{hsq00,dl09,gds+10}), and photoionization and/or mechanical feedback from starbursts or AGN (e.g. \citealt{ts00,gal+09}).  Supporting evidence for the cooling flow scenario comes from those blobs lacking any visible power source, in both the optical and infrared \citep{nfm+06,sjlm08}, while the significant number of blobs containing massive, luminous galaxies or AGN suggests that internal heating plays an important role \citep{gal+09,cst+11}.  

Most recently, attention has been directed to the effects of \lya\ radiative transfer and resonant scattering, suggesting that cooling alone is insufficient to account for the observed luminosity \citep{fkd+10}.  Also recently, stacking of narrowband \lya\ images of galaxies at \ztwo--3 has revealed extended, diffuse \lya\ halos, suggesting that such emission is a general property of high redshift galaxies, and that the distinction between a galaxy and a \lya\ blob may simply be a matter of the surface brightness threshold of the observation \citep{sbs+11}.   The typical sizes of these extended halos ($\sim80$ kpc) are comparable in extent to the metals and \ion{H}{1} detected in absorption near galaxies \citep{ses+10}, implying that the \lya\ emission is due to resonant scattering of photons produced in galactic \ion{H}{2} regions by outflowing gas.  Similarly, \citet{zcw+10} predict that resonant scattering will produce extended \lya\ halos around galaxies at high redshift, resulting in a coupling between the observed \lya\ emission and the properties of the circumgalactic and intergalactic media on larger scales.

In this Letter we present six large \lya\ blobs discovered in the course of a narrowband imaging survey for \lya\ emission associated with a known protocluster.   The $z=2.300 \pm 0.015$ HS 1700+643 protocluster  \citep{sas+05} has a redshift-space galaxy overdensity $\delta \sim 7$ and a mass scale of $\sim1.4\times10^{15}$ \msun, indicating that it will evolve into a rich galaxy cluster by $z=0$.  Galaxies in the overdensity have typical masses and ages $\sim2$ times larger than comparably selected field galaxies, while an analysis of {\it HST} ACS images finds no significant morphological differences between protocluster and field galaxies \citep{psl+07}.  The protocluster has also been the target of a 200-ks {\it Chandra}/ACIS-I observation, which indicated an excess of AGN activity relative to a similarly selected field sample \citep{dnl+10}.

We use a cosmology with $H_0=70\;{\rm km}\;{\rm s}^{-1}\;{\rm Mpc}^{-1}$, $\Omega_m=0.3$, and $\Omega_{\Lambda}=0.7$.

\section{Data}
\label{sec:data}
The \lya\ blobs were detected in a deep, narrowband image centered at 4010 \AA, using the LFC Wide-Field Imager on the Hale 200-inch telescope at Palomar Observatory and a custom filter with $\rm{FWHM}=90$ \AA\ designed to match \lya\ emission at $z=2.3$.  The 22.3 hr exposure has a 3$\sigma$ depth of $NB\sim27.8$ mag arcsec$^{-2}$ and an area of 219 arcmin$^{2}$.  The observations, data reduction and resulting catalog of \lya\ emitters are described in detail elsewhere (Erb et al., in prep).  In brief, individual exposures were 1800 s, with sky subtraction using the IRAF task ``imsurfit" and masking of satellite tracks done on individual frames.  The individual frames were then combined and matched to a continuum image created via a linear combination of $U_n$ and $G$ images (the peak transmission of the 4010 \AA\ filter lies directly between the $U_n$ and $G$ bands).  Final background subtraction and object detection was done with SExtractor, and the resulting final image and background image were carefully examined for residual artifacts which could affect the blob profiles.  Spectroscopic followup of portions of the blobs was carried out with the Low Resolution Imaging Spectrometer (LRIS) on the Keck I telescope, following procedures described by \citet{ssp+04}.

The HS 1700+64 field has been the subject of extensive deep, multiwavelength imaging: near-IR $J$ and $K_s$ observations are presented by \citet{ess+06mass}, {\it Spitzer} IRAC observations in all four channels by \citet{bhf+04} and \citet{sse+05}, and 24 \micron\ {\it Spitzer} MIPS observations by \citet{rep+10}.  This imaging covers a smaller area than the optical and narrowband images; two of the blobs presented here (Blobs 5 and 6) are not covered by the IR observations.

We identify rest-frame UV-selected protocluster galaxies and candidates as described by \citet{ssp+04,sas+05}, and select Distant Red Galaxies (DRGs) with $J-K_s>2.3$ \citep{flr+03}.  The DRGs have a broad redshift range, from $z\sim1.2$ to $z\sim3.7$ with a median of $z\sim2.7$, though large spectroscopic samples have been difficult to obtain because of their optical faintness \citep{wvf+09}.  Additional followup of the spectroscopic sample indicates that the DRGs are massive and divided between dusty galaxies with high star formation rates and compact, quiescent galaxies \citep{kvf+09}.  The catalog of DRGs is largely complementary to the UV-selected galaxy catalog, with relatively small overlap of $\sim12$\% between the two selection methods \citep{res+05}.  

\begin{deluxetable*}{l l l c c l l l c c}
\tablewidth{0pt}
\tabletypesize{\footnotesize}
\tablecaption{\lya\ Blobs in the HS 1700+64 Protocluster at $z=2.3$\label{tab:blobs}}
\tablehead{
\colhead{ID} & 
\colhead{RA} & 
\colhead{Dec} & 
\colhead{Area\tablenotemark{a}} &
\colhead{$a$\tablenotemark{b}} &
\colhead{$e$\tablenotemark{c}} &
\colhead{PA\tablenotemark{d}} &
\colhead{$W_0$\tablenotemark{e}} &
\colhead{$F_{\lya}$\tablenotemark{f}} &
\colhead{$L_{\lya}$} \\
\colhead{} &
\colhead{(J2000)} &
\colhead{(J2000)} &
\colhead{(arcsec$^2$)} &
\colhead{(kpc)} &
\colhead{} &
\colhead{(deg)} &
\colhead{(\AA)} &
\colhead{($10^{-16}$ erg s$^{-1}$ cm$^{-2}$)} &
\colhead{($10^{43}$ erg s$^{-1}$)}
}
\startdata

BLOB 1 & 17:00:46.99 & +64:15:43.5 & 105 & 151 & 0.50 & \phantom{0}97 & 126 & 4.6 & 1.8\\
BLOB 2 & 17:00:56.56 & +64:13:27.2 & \phantom{0}61 & 149 & 0.56 & 156 & 125 & 3.5 & 1.4\\
BLOB 3 & 17:01:04.16 & +64:12:11.8 & 128 & 125 & 0.12 & 140 & 129 & 6.6 & 2.6\\
BLOB 4 & 17:01:32.53 & +64:14:16.4 & \phantom{0}95 & 137 & 0.22 & 110 & 158 & 5.6 & 2.2\\
BLOB 5 & 17:01:41.31 & +64:14:03.3 & \phantom{0}53 & 103 & 0.19 & \phantom{0}99 & \phantom{0}71 & 3.2 & 1.2\\
BLOB 6 & 17:01:54.54 & +64:13:39.6 & \phantom{0}78 & 119 & 0.17 & \phantom{0}\phantom{0}9 & 121 & 3.5 & 1.4

\enddata
\tablenotetext{a}{Area of emission 1$\sigma$ above sky in
  continuum-subtracted narrowband image.}
\tablenotetext{b}{Major axis diameter of ellipse fit to $1.5 \times
  10^{-18}$ erg s$^{-1}$ cm$^{-2}$ arcsec$^{-2}$ contour.}
\tablenotetext{c}{Ellipticity $e=1-b/a$ of best-fit ellipse.}
\tablenotetext{d}{Position angle of best-fit ellipse.}
\tablenotetext{e}{Rest-frame equivalent width of \lya\ emission.}
\tablenotetext{f}{Flux within area given in column 4.}

\end{deluxetable*}

\section{Six \lya\ Blobs at $z=2.3$}
\label{sec:blobfacts}
The narrowband image contains six regions of extended \lya\ emission, classified as ``blobs" based on an isophotal size of at least 50 arcsec$^{2}$ with a 1$\sigma$ surface brightness limit of $NB=29$ mag arcsec$^{-2}$ or $1.5\times10^{-18}$ erg s$^{-1}$ cm$^{-2}$ arcsec$^{-2}$.  With this threshold, the blobs have areas of $\sim50$--130 arcsec$^{2}$ and luminosities of $\sim10^{43}$ erg s$^{-1}$.  Because of the irregular shape of the blobs, linear size measurements are difficult.  In order to estimate the blobs' linear extent, elongation and position angle, we fit an ellipse to the largest $1.5\times10^{-18}$ erg s$^{-1}$ cm$^{-2}$ arcsec$^{-2}$ contour.  The major axis of the ellipse then approximates the size of the blob, and the ellipticity and position angle measure its elongation and orientation.   Properties of the blobs are given in Table \ref{tab:blobs}, and images of the blobs with overlaid contours and the best-fit ellipses are shown in Figure \ref{fig:blobpics}.   

\subsection{Space Density}
Given the 219 arcmin$^2$ area and filter bandwidth sensitive to \lya\ emission from $z=2.24$ to $z=2.36$, the effective survey volume is $V\simeq10^5$ comoving Mpc$^3$.  We therefore estimate the  space density of blobs in the protocluster to be $n\sim6\times10^{-5}$ Mpc$^{-3}$.   In a blind, field survey at the same redshift and to a similar luminosity threshold (though allowing a smaller total blob area), \citet{yzt+09} found a blob density of $2.5\times10^{-6}$ Mpc$^{-3}$, a factor of 24 lower than the density in the HS 1700+64 protocluster.  Even when accounting for the protocluster's $\delta \sim 7$ galaxy density enhancement, the blob density remains higher by a factor of $\sim3$.  \citet{yzt+09} reached a similar conclusion when comparing their observed blob density with that of the $z=3.1$ overdensity in SSA22 \citep{myh+04}.  Our results therefore confirm earlier conclusions that \lya\ blobs are more likely to be found in significantly overdense environments. 

\subsection{Redshifts}
\label{sec:redshifts}
All of the blobs have been observed spectroscopically, several at more than one position, and confirmed to be associated with the $z=2.3$ protocluster.  \lya\ emission redshifts range from $z=2.267$ (Blob 5) to $z=2.3082$ (Blob 2), corresponding to a velocity range of $\sim3700$ \kms\ ($\sim1800$ \kms\ if Blob 5 is not included).  There is no trend of redshift with blob position, and the range of blob redshifts is somewhat wider than the redshift range of the continuum-selected galaxy overdensity, $z=2.300 \pm 0.015$ \citep{sas+05}.  

\begin{figure*}[htbp]
\plotone{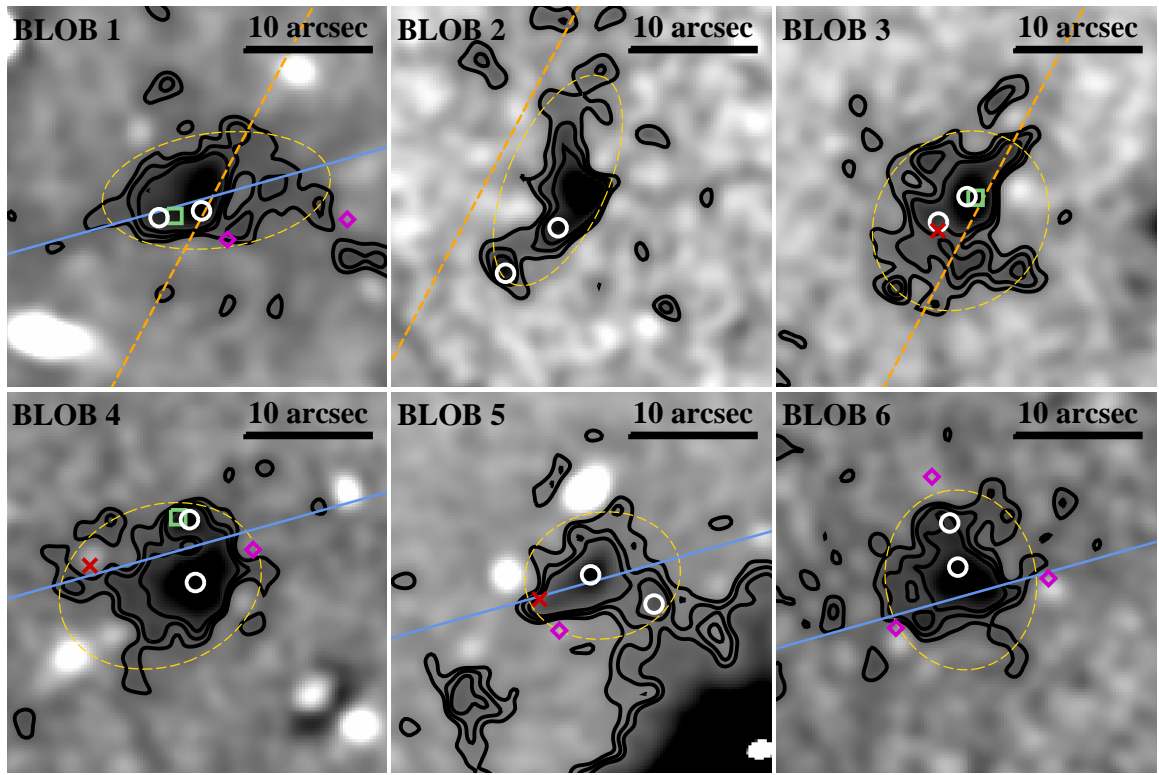}
\caption{The six \lya\ blobs are marked with contours on the smoothed, continuum-subtracted narrowband image.  The contours are created after smoothing the image with a Gaussian of $\rm{FWHM}=7$ pixels ($\simeq1.7\arcsec$), and are marked at surface brightness levels of 1.5, 3 and $4.5\times10^{-18}$ erg s$^{-1}$ cm$^{-2}$ arcsec$^{-2}$.  Dashed yellow lines show ellipses fit to the largest $1.5\times10^{-18}$ erg s$^{-1}$ cm$^{-2}$ arcsec$^{-2}$ contour.  White circles show portions of the blobs classified as narrowband \lya\ emitters, red crosses show UV continuum-selected galaxies spectroscopically confirmed to belong to the $z=2.3$ protocluster, magenta diamonds show UV-selected \ztwo\ candidates with unknown redshifts, and green squares show DRGs whose redshifts are unknown (except the DRG associated with Blob 3, which has an absorption redshift associating it with the blob).  The solid blue line is a least-squares fit to the positions of blobs 1, 4, 5 and 6 (also shown as the dashed line in Figure \ref{fig:blobmap}), and the dashed orange line is fit to the positions of blobs 1, 2 and 3 (also shown as the dot-dashed line in Figure \ref{fig:blobmap}).  The images are oriented with north up and east to the left, and the scale  bar in each window corresponds to 10 arcsec, or 82 proper kpc at $z=2.3$.  The bright object in the lower right corner of the image of blob 5 is a foreground star.}
\label{fig:blobpics}
\end{figure*}

\subsection{Associations with Continuum-Selected Galaxies and Blob Kinematics}
Most of the blobs are associated with continuum-selected galaxies confirmed or likely to lie in the protocluster; however, none are associated with known AGN, and none are detected in a deep {\it Chandra}/ACIS-I observation of the field \citep{dnl+10}.  In this section we briefly describe the galaxies associated with each blob, and summarize what is known about the blobs' kinematics.

{\it Blob 1}.  The bright ($K_s=19.2$), near-IR selected galaxy DRG64 lies near the center of Blob 1, between the two portions of the blob selected as \lya\ emitters.  DRG64 is also a bright 24 \micron\ source.  The redshift of this galaxy is unknown, but given the low density of DRG candidates in the field the probability that this is a chance association is low.  \lya\ spectroscopy of the blob reveals extended emission at $z=2.2929$, with velocity width $\sigma \sim 750$ \kms\ and no significant velocity shift over the spatial extent of the emission.

{\it Blob 2}.  Blob 2 is the only one of the six blobs not coincident with any UV- or IR-selected galaxy candidates.  It is also the blob with the largest observed velocity gradient: spectroscopy of \lya\ emission from three different portions of the blob reveals emission at $z=2.3082$ at the core of the blob, $z=2.3065$ at the northern of the two points marked as narrowband-selected \lya\ emitters in Figure \ref{fig:blobpics}, and $z=2.2930$ at the southern edge of the blob.  \lya\ emission from the edge of the blob is therefore blueshifted by $\sim 1400$ \kms\ with respect to emission from the core of the blob 70 kpc away.  

{\it Blob 3}.  Blob 3 contains the UV-selected galaxy MD109, with a redshift from \Ha\ emission of $z=2.2942$ \citep{ess+06mass}, and the IR-selected galaxy DRG38.  DRG38 is coincident with a portion of the blob selected as a \lya\ emitter, and followup spectroscopy reveals weak interstellar absorption lines with a redshift $z=2.260$ as well as \lya\ emission at $z=2.290$.  \lya\ emission from the core of the blob is broad and multi-peaked, with $\sigma \sim 1400$ \kms.  

{\it Blob 4}.  This blob contains the IR-selected DRG72, with unknown redshift, and the UV-selected galaxy BX909, a strong 24 \micron\ source with $z_{\rm abs}=2.291$ and \Ha\ redshift $z=2.2937$.  An additional UV-selected source with unknown redshift, BX912, lies near the western edge of the blob.  Spectroscopy of the two portions of the blob marked with white circles in Figure \ref{fig:blobpics} yields \lya\ redshifts of $z=2.296$ at the core and $z=2.288$ at the northern edge, for a velocity difference of $\sim 700$ \kms.

{\it Blob 5}.  Blob 5 contains the UV-selected galaxy BX888, with $z_{\lya}=2.270$ and $z_{\rm abs}=2.263$, and a UV-selected galaxy of unknown redshift lies near the southern edge of the blob.  As noted above, Blob 5 is at slightly lower redshift than the other blobs.  The core of the blob shows a double-peaked \lya\ emission line with $z=2.267$ and $z=2.272$, for a difference between the peaks of $\sim450$ \kms.  The spectrum may also show a weak \ion{C}{4} emission line.  Blob 5 is not covered by IR imaging.

{\it Blob 6}.  Blob 6 contains no galaxies with known redshifts, but three UV-selected \ztwo\ candidates lie just beyond the edges of the blob.  \lya\ emission is spatially extended with $z=2.299$ and no velocity gradient.  Blob 6 is not covered by IR imaging.

\begin{figure*}[htbp]
\plotone{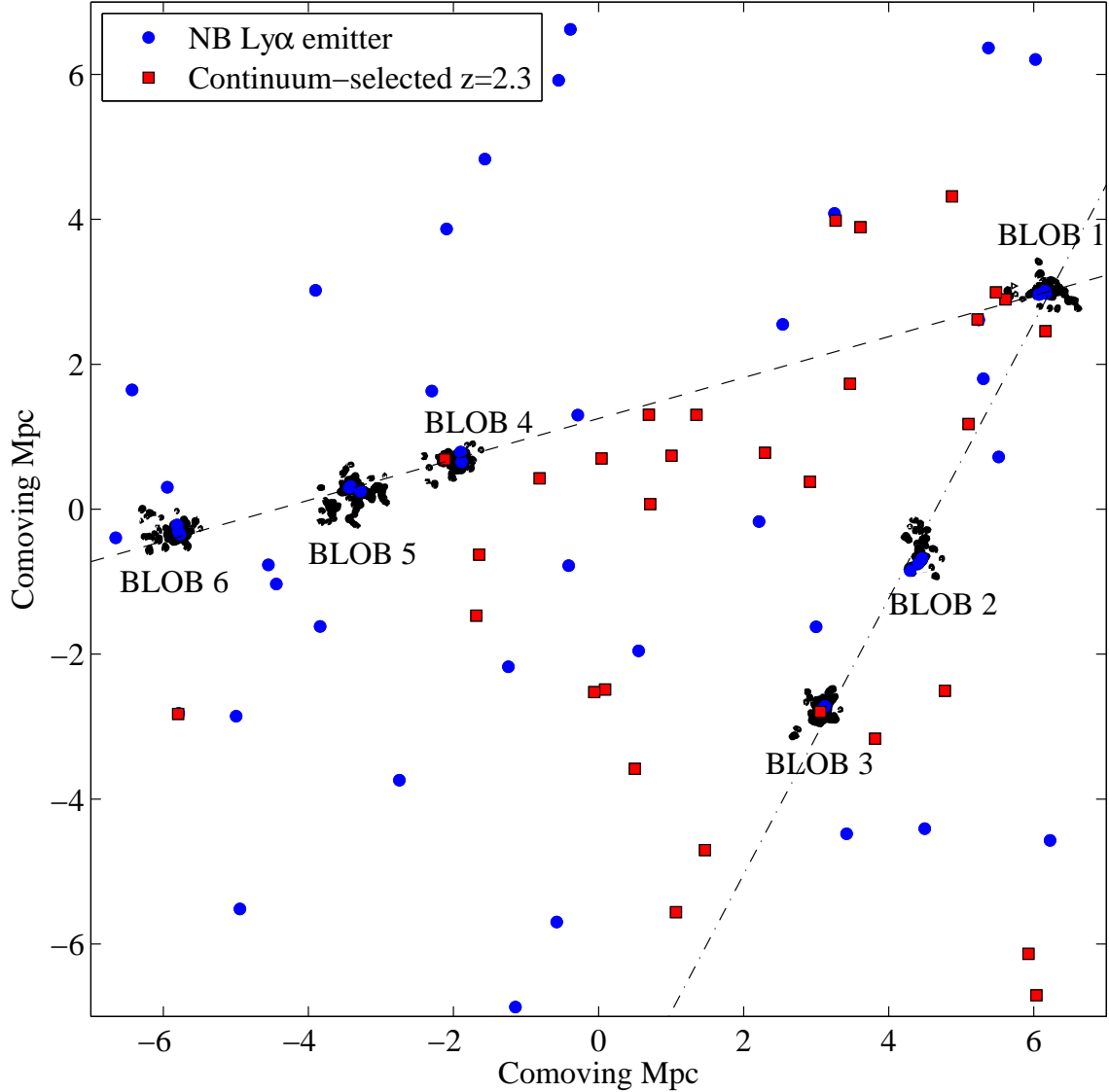}
\caption{The six \lya\ blobs, NB-selected \lya\ emitters, and rest-frame UV continuum-selected members of the $z=2.3$ protocluster.  The dashed line is a least-squares fit to the positions of blobs 1, 4, 5 and 6, and the dot-dashed line is a fit to the positions of blobs 1, 2 and 3.}
\label{fig:blobmap}
\end{figure*}

\section{Filamentary Structure and Blob Alignment}
\label{sec:lss}
The six \lya\ blobs span a region of approximately $12\times 6$ comoving Mpc, and are distributed along two lines which intersect at the position of Blob 1 (see Figure \ref{fig:blobmap}; the edge of the narrowband image is just to the left of Blob 6, so the structure may extend farther to the east).  Rest-frame UV-selected galaxies spectroscopically confirmed to be associated with the protocluster also tend to lie along and between these two lines, while narrowband-selected $z\sim2.3$ \lya\ emitters are more widely and randomly distributed. 

The blobs therefore define two filamentary structures, viewed in projection.  As noted in Section \ref{sec:redshifts}, the \lya\ redshifts of the six blobs correspond to a velocity range of $\sim3700$ \kms, with no correlation between position and redshift.  Without systemic redshifts for the blobs, we cannot determine how much of these velocity differences are due to outflows or inflows within the blobs, relative velocities between the blobs, or real differences in physical distance.

We quantify the alignment of the blobs by fitting lines to the positions of Blobs 1, 4, 5 and 6, and to the positions of Blobs 1, 2 and 3.  These two lines are shown in Figure \ref{fig:blobmap}, by the dashed and dot-dashed lines respectively.  They are also shown in Figure \ref{fig:blobpics}, where the Blob 1-4-5-6 line (hereafter Line 1456) is shown in solid blue, and the Blob 1-2-3 line (hereafter Line 123) is dashed orange.  Inspection of Figure \ref{fig:blobpics} shows that these lines pass through or very close to all six blobs, the intersection of the two lines is at the center of Blob 1, and that all of the galaxies confirmed to be associated with the blobs, and all three of the DRGs, lie on or within only a few arcseconds of the lines.

We also compare the angles of these lines to the orientation of the blobs.  Line 1456 is at an angle of 106$^{\circ}$ (measured east from north as usual), while the angle of Line 123 is 152$^{\circ}$.  We determine the position angle of each blob by fitting an ellipse to the largest contour at the $1.5\times10^{-18}$ erg s$^{-1}$ cm$^{-2}$ arcsec$^{-2}$ flux level.  The ellipses are shown by the dashed yellow lines in Figure \ref{fig:blobpics}, and the  resulting angles are listed in Table \ref{tab:blobs}.  We find that the position angles of Blobs 1, 4, 5 and 6 are 97$^{\circ}$, 110$^{\circ}$, 100$^{\circ}$ and 9$^{\circ}$ respectively.  With the exception of Blob 6, all of these blobs are aligned to within less than 10$^{\circ}$ along the line connecting them.  Similarly, we measure position angles of 156$^{\circ}$ and 140$^{\circ}$ for Blobs 2 and 3 respectively, for offsets of 4$^{\circ}$ and 12$^{\circ}$ with respect to Line 123.  Thus five of the six blobs are aligned to within $\sim10$$^{\circ}$ of the lines connecting them.  We also note that the position angle of Blob 6, the only blob that is misaligned, is significantly affected by the narrow extended tail on the southwest edge of the blob; without this tail, the remainder of the blob is roughly aligned with Line 1456. 

Given the nebulous, irregular borders of the blobs, the difficulty of reliably assigning an orientation, and the fact that only two show significant ellipticity $e\geq0.5$, we should view this result with some caution.  Nevertheless, the probability of the observed alignment happening by chance is extremely low.  Assigning position angles at random to each of the six blobs and repeating the process $10^6$ times, we find that at least five blobs are aligned to within 12$^{\circ}$ of the lines connecting them only 0.04\% of the time.  This suggests that the structure of the blobs is indeed related to the larger scale structure of which they are a part.

\section{Summary and Discussion}
\label{sec:disc}
We have presented six $\sim100$ kpc \lya\ blobs in a protocluster at $z=2.3$, the richest field of giant \lya\ blobs detected to date.  These blobs precisely define two intersecting filaments, much like the filaments along which galaxies form in cosmological simulations of structure formation (e.g.\ \citealt{swj+05}).  Moreover, we have shown that five of the six blobs are morphologically aligned with these filaments, indicating a connection between the processes responsible for \lya\ emission in the blobs and the environment on larger scales.

Previous observations have firmly established the connection between \lya\ blobs and environmental overdensities; as just one particularly relevant example, \citet{myh+05} used a wide-field survey of \lya\ emitters in the SSA22 field to establish the existence of large-scale filamentary structure at $z=3.1$, and showed that the two giant \lya\ blobs in that field are associated with those filaments, confirming the association of blobs with density peaks within overdense regions \citep{sas+00}.  More recently, \citet{myh+11} suggested a connection between the morphology of giant \lya\ blobs and environment, finding that filamentary, elongated blobs reside in average density environments (as defined by the overdensity of \lya\ emitters), while more circular blobs are found in both average and overdense regions.  This observation  leads to the hypothesis that circular blobs are associated with outflows from massive galaxies, while more filamentary blobs may be related to the accretion of cold gas from the IGM.  The blobs presented here somewhat complicate this scenario, as they show a variety of morphologies from more to less elongated but are all found in relatively close proximity and in an overdense environment.  On the other hand, the most elongated blob, Blob 2, is also the best candidate for emission arising from cold accretion, since it is the only one of the six blobs containing no UV or IR-selected galaxy candidates.  Blob 2 also has the largest observed velocity gradient among the six, at $\sim1400$  \kms\ over 70 kpc, though this is difficult to interpret in terms of infall or outflow without systemic redshifts for the emitting regions.

The two new pieces this work brings to the picture are the nearly perfectly linear filaments on which the six $z=2.3$ blobs fall, and the general alignment of the blobs with those structures.  Perhaps the most obvious interpretation of this result is that at least some of the \lya\ emission from the blobs arises from gas falling into massive galaxies forming along the filaments.  This may be a reasonable conclusion; although recent simulations suggest that emission from cold accretion is insufficient to entirely power the large \lya\ blobs, it may play an important role even if not energetically dominant \citep{fkd+10}.  However, most of the blobs have galaxies associated with them, and three of those galaxies are IR-selected DRGs; the DRGs are likely to be massive, and given their low surface density, their association with the blobs is unlikely to be coincidental.  This fact, coupled with the large spatial and velocity extent of gas outflowing from galaxies and the ability of \lya\ photons from within the galaxies to resonantly scatter in that gas \citep{ses+10,sbs+11,zcw+10}, suggests that heating from within is also likely to be a significant power source for the blobs.   Resonant scattering may also result in \lya\ emission from filamentary gas around galaxies whether or not the gas is accreting, complicating the relationship between blob power sources and morphologies.

The \lya\ blobs in the HS 1700+64 protocluster suggest a complex interplay between outflowing and infalling gas, mediated by the resonant scattering of \lya\ photons which couples the emission from galaxies to the environment on larger scales.  We look forward to much deeper narrowband images of this region, when the technology becomes available; such images will likely reveal more extended \lya\ emission, possibly further tracing the filamentary structures identified by the blobs.  Additional observations will also more thoroughly reveal the velocity structure of the blobs, for better constraints on their emission mechanisms and on the complex processes of radiative transfer  which produce them.

\acknowledgements{Alice Shapley, Naveen Reddy and Max Pettini provided essential contributions to the survey on which this work is based.  
We also thank the staffs of Palomar and Keck Observatories for assistance with the observations, Yuichi Matsuda for useful suggestions, and the referee for helpful comments.  D.K.E. acknowledges support from the University of Wisconsin Research Growth Initiative, and C.C.S. from the US National Science Foundation through grants AST-0606912 and AST-0908805, with additional support from the John D. and Catherine T. MacArthur Foundation and the Peter and Patricia Gruber Foundation.

\bibliographystyle{apj}

\end{document}